\documentclass{article}
\usepackage{graphicx}

\begin{document}

\begin{center}
 {\bf The Impact of Type Ia Supernova Explosions on their Companions in Binary System}
\end{center}
\begin{center}
X. Meng$^{\rm 1, 2}$, X. Chen$^{\rm 1}$ and Z. Han$^{\rm 1}$
\end{center}
\begin{center}
$^{\rm 1}$National Astronomical Observatories/Yunnan Observatory,
the Chinese Academy of Sciences, Kunming, 650011, China,
conson859@msn.com
\end{center}
\begin{center}
$^2$ Graduate School of the Chinese Academy of Sciences
\end{center}

\begin{abstract}
Using a simple analytic method, we calculate the impact effect
between the ejecta of a SN Ia and its companion to survey the
influence of initial parameters of the progenitor's system, which
is useful for searching the companion in a explosion remnant. The
companion models are obtained from Eggleton's evolution code.The
results are divided into two groups based on mass transfer stage.
For a given condition, more hydrogen-rich material is stripped
from the envelope of a Hertzsprung-gap companion than that of a
main-sequence companion, while a larger kick velocity and a larger
luminosity are gained for a main-sequence companion. The kick
velocity is too low to significantly affect the final spatial
velocity of the companion, which is mainly affected by the initial
parameters of the progenitor systems. The spatial velocity of the
stripped material has an upper limit within the range of 8000 -
9500 km/s, which only depends on the total kinetic energy of the
explosion.The stripped mass, the ratio of the stripped mass to the
companion mass and the kick velocity of the companion all
significantly depend on the initial companion mass and orbital
period. Our model may naturally explain the spatial velocity of
the star G in the remnant of Tycho's supernova, while an
energy-loss mechanism is needed to interpret its luminosity.
\end{abstract}

\section{Introduction}\label{sect:1}
Type Ia supernova (SNe Ia) have been successfully used to
determine cosmological parameters, e.g. $\Omega_{\rm M}$ and
$\Omega_{\Lambda}$ (Reiss et al. \cite{REI98}; Perlmutter et al.
\cite{PER99}), although we do not know about the exact nature of
SNe Ia, especially about their progenitors. The most widely
accepted model is a single degenerate Chandrasekhar mass model, in
which a carbon-oxygen white dwarf (CO WD) increases its mass by
accreting hydrogen- or helium-rich matter from its companion, and
explodes when its mass approaches the Chandrasekhar mass limit
(Whelan \& Iben \cite{WI73}). The companion may be a main-sequence
star (WD+MS) or a red-giant star(WD+RG) (Yungelson et al.
\cite{YUN95}; Li et al. \cite{LI97}; Hachisu et al. \cite{HAC99a}
\cite{HAC99b}; Nomoto et al. \cite{NOM99}; Langer et al.
\cite{LAN00}). Hachisu \& Kato (\cite{HK03a}, \cite{HK03b}) argued
that supersoft X-ray sources, which belong to the WD+MS channel,
may be good candidates for the progenitors of SNe Ia. Observation
of the remnant of SN 1572 (Tycho's supernova) favors the model of
WD+MS and suggests that a star named star G is likely to be the
companion of Tycho's supernova ( Ruiz-Lapuente et al.
\cite{RUI04}; Branch \cite{BRA04}).

In the single degenerate model, the supernova ejecta collides into
the envelope of its companion and strips some hydrogen-rich
material from the surface of the companion (Cheng \cite{CHE74};
Wheeler et al. \cite{WHE75}; Fryxell \& Arnett \cite{FRY81}; Taam
\& Fryxell \cite{TAA84}; Chugai \cite{CHU86}; Livne et al.
\cite{LIV92}; Langer et al. \cite{LAN00}). The stripped
hydrogen-rich material may reveal itself by narrow H$_{\rm
\alpha}$ emission or absorption lines in later-time spectra of SNe
Ia (Chugai \cite{CHU86}; Filippenko \cite{FIL97}). Marietta et al.
(\cite{MAR00}, hereafter M00) ran several high-resolution
two-dimensional numerical simulations of the collision between the
ejecta and the companion, which is a MS star, a subgiant (SG) star
or a red giant (RG) star. They found that about 0.15 $M_{\rm
\odot}$ - 0.17 $M_{\rm \odot}$ of hydrogen-rich material is
stripped from a MS or a SG companion and there is no difference
between the two companions. After the impact, the companion gains
a small kick velocity and its luminosity will rise dramatically to
as high as 5000 $L_{\rm \odot}$. However, the SG companion model
in M00 was gained by adjusting the entropy profile of the
companion to simulate the effect of binary mass transfer and the
MS companion model in M00 was represented by a 1.0 $M_{\odot}$
solar model. Their companion models were not from a detailed
binary evolution calculation and the study was only for $Z=0.02$,
which lead the results to be different from an actual case. In
this paper, we use some companion models obtained from the
Eggleton's evolution code (\cite{EGG71}, \cite{EGG72},
\cite{EGG73}), which are more realistic than that in M00, to
examine the effects of some initial parameters on the collision by
a simple analytic method.

\section{Method and Results}\label{sect:2}
\subsection{method}\label{subs:2.1}
\begin{figure}
   \vspace{2mm}
   \begin{center}
\includegraphics[width=3in, angle=-90]{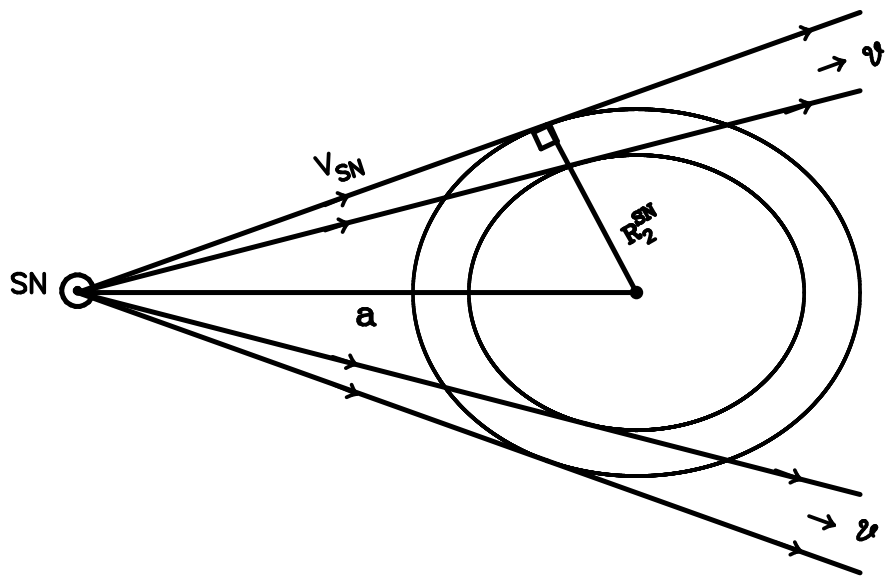}
   \caption{Schematic for the impact between the supernova's ejecta and its companion.
    Supernova ejecta collides into the envelope of its companion and strips some hydrogen-rich
   material from the surface of the companion.}
   \label{Fig1}
   \end{center}
\end{figure}

We consider the case where a CO WD accretes matter from its
companion which may be a MS star or a Hertzsprung-gap (HG) star.
When the CO WD increases its mass to close to the Chandrasekhar
mass, i.e. 1.378 $M_{\odot}$ (Nomoto, Thielemann \& Yokoi
\cite{NTY84}), it explodes as a SN Ia. Using the method of Han \&
Podsiadlowski (\cite{HAN04}), we get 23 companion models for
different metallicities which are listed in table \ref{Tab:1}.
Then, the changes in the secondary structure due to mass transfer
are taken into account naturally. An optically thick wind (Hachisu
et al. \cite{HAC96}) is used to calculate the mass loss and
angular moment loss from the binary system. The prescription of
Hachisu (\cite{HAC99a}) about hydrogen accretion is adopted to
calculate the growth of the WD mass. The mass accumulation
efficiency for helium-shell flashes is from Kato \& Hachisu
(\cite{KH04}). We changed one initial parameter and fixed the
others to test the effect of different parameters on the final
results. In table \ref{Tab:1}, we see that the mass transfer
between a CO WD and its companion may begin as the companion is a
MS star or a HG star. Note that the definition of HG stars in this
paper is similar to that of the SG model in M00. Evolving these
binaries, we get the companion models as the WD mass increases to
1.378 $M_{\odot}$. After the explosion, a large amount of material
is ejected as a series of spherically expanding shells and impact
on the surface of the companion. The leading edge of these
expanding shells collides into the envelope of the companion with
a velocity $V_{\rm SN,0}$ at $t_{\rm 0}=a/V_{\rm SN,0}$, where $a$
is the orbital separation of the binary system at the moment of
the explosion and it is deduced from Eggleton's equation (Eggleton
\cite{EGG83}) by assuming that the companion radius $R_{\rm
2}^{\rm SN}$ equals the critical radius of its Roche lobe $R_{\rm
cr}$. We assume that the density in each spherical shell is
uniform and that each shell moves at a fixed velocity $V_{\rm
SN}=a/t$, where $t$ and $t_{\rm 0}$ both take the moment of the
explosion as the zero point of time. The density of the expanding
shell at a distance $r=a$ from the explosion center is scaled as
  \begin{equation}
    \rho_{\rm SN}=\frac{3M_{\rm SN}}{4\pi a^{3}}(\frac{t_{\rm 0}}{t})^{3},
  \end{equation}
after $t\geq t_{\rm 0}$, where $M_{\rm SN}$ is the total mass of
the CO WD at explosion, i.e. 1.378 $M_{\odot}$ (Chugai
\cite{CHU86}). The definitions of the density and the velocity are
similar to those of M00. Then, the total kinetic energy of the
ejecta is
  \begin{equation}
 E_{\rm k}=\int_{t_{\rm 0}}^{\rm \infty}\frac{1}{2}\rho _{\rm SN}\cdot V\cdot
dt\cdot 4\pi a^{\rm 2}\cdot V^{\rm 2}=\frac{3}{10}M_{\rm SN}V_{\rm
SN,0}^{\rm 2}
  \end{equation}
and the total momentum of the ejecta is
  \begin{equation}
 P_{\rm t}=\int_{t_{\rm
0}}^{\infty}\rho _{\rm SN}\cdot V\cdot dt\cdot 4\pi a^{\rm 2}\cdot
V=\frac{3}{4}M_{\rm SN}V_{\rm SN,0},
  \end{equation}
 where $M_{\rm SN}$ is the total mass of the ejecta. As shown
in Fig. \ref{Fig1}, the ejecta mass which collides into the $i$th
slab in the envelope is calculated by
  \begin{equation}
M_{\rm i}^{\rm SN}= M_{\rm SN}\cdot \frac{R_{\rm 2, i}^{\rm
2}-R_{\rm 2, i-1}^{\rm 2}}{4a^{\rm 2}},
  \end{equation}
where $R_{\rm 2, i}$ is the radius of the $i$th slab stripped from
the companion.  Then, the momentum of $M_{\rm i}^{\rm SN}$ is
$P_{\rm i}=\frac{3}{4}M_{\rm i}^{\rm SN}V_{\rm SN,0}.$ Assuming
that the ejecta and the envelope material leave with the same
velocity $v$ along the same direction of the ejecta velocity, we
may get $v$ by momentum conservation. If $v$ exceeds the escape
velocity $V_{\rm esc}$ of the companion, the envelope material is
stripped.Then, the amount of the stripped material is the sum of
all the material in these stripped slabs. Since only the kinetic
properties of the ejecta are considered, the composition of the
ejecta is not considered. After the impact of the ejecta, a shock
like a bowl develops (Fryxell \& Arnett \cite{FRY81}; M00).
However, our method is unable to calculate the effect of the
shock. We discuss whether our simplification is reasonable or not
in the next subsection. The kinetic energy of the supernov ejecta
is assumed to be $1\times 10^{\rm 51}$ erg, which corresponds to
the lower limit of the kinetic energy of normal SNe Ia (Gamezo et
al. \cite{GAM03}).

If the supernova ejecta injected into the companion envelope can
not strip the material from the surface of the companion, i.e.
$v<V_{\rm esc}$, the ejecta will settle in the companion and the
momentum of the ejecta transfers to the companion. The companion
then gains a kick velocity $V_{\rm kick}$ (Cheng \cite{CHE74};
M00). During this process, some material reverses to the explosion
center and the companion gains an added momentum (Fryxell \&
Arnett \cite{FRY81}; M00). We neglect this effect because it does
not significantly affect the final results (Fryxell \& Arnett
\cite{FRY81}; M00). The kick velocity is gained by the
conservation of linear momentum. Note that the ejecta is not
always parallel to the axis between explosion center and companion
center. However, we take the momentum of the ejecta settled in the
companion as the ejecta's momentum paralleled to the axis and
neglected the effect of angle on the momentum paralleled to axis
because the angle is very small.
\subsection{discussion of the method}\label{subs:2.2}

\begin{figure}
   \vspace{2mm}
   \begin{center}
\includegraphics[width=3in, angle=-90]{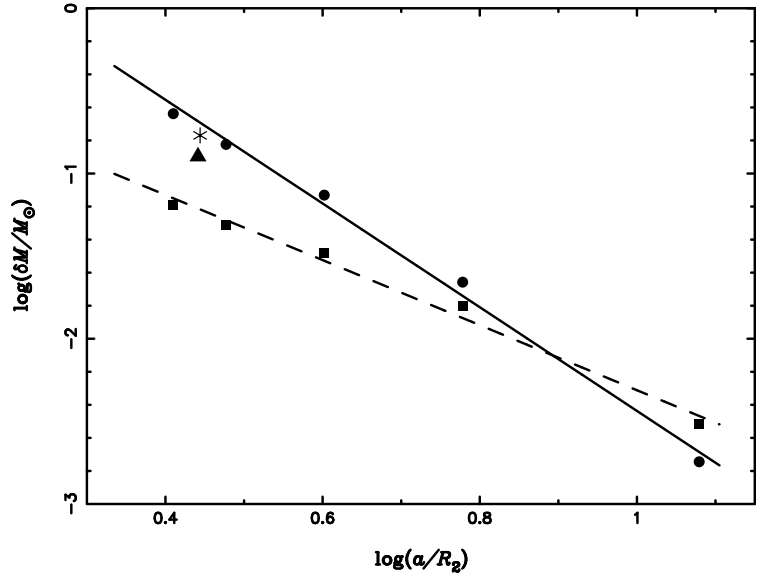}
   \caption{Comparison between the stripped masses in this paper and those of M00. Filled circles are from M00
   and filled squares are our
   results for the MS model. Dashed line and solid line fit linearly our results of MS models and those of M00, respectively.
   The triangular and Hexagonal points are our result for the SG model and that of M00, respectively.}
   \label{Fig2}
   \end{center}
\end{figure}

\begin{figure}
   \vspace{2mm}
   \begin{center}
\includegraphics[width=3in, angle=-90]{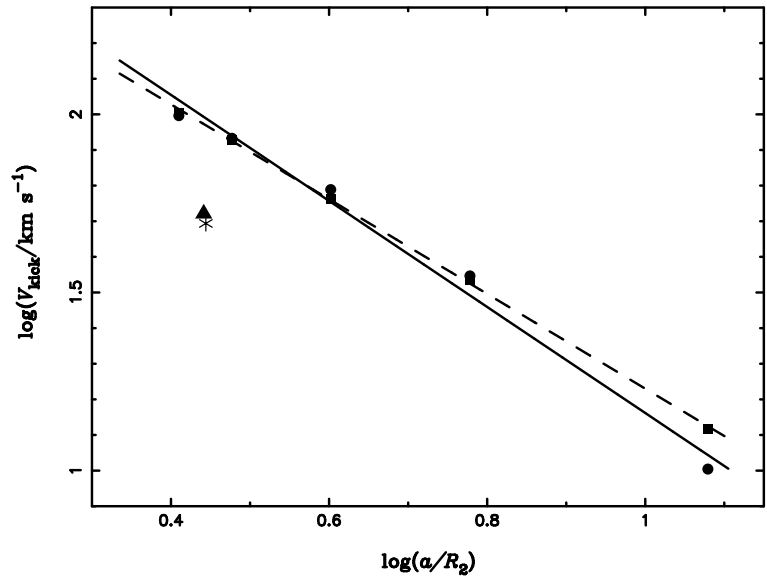}
   \caption{Similar to Fig. \ref{Fig2}, but for kick velocity.}
   \label{Fig3}
   \end{center}
\end{figure}

It is well known that a shock will develop after the impact of the
ejecta. A large part of the material in the companion's envelope
will be heated by the shock and then be vaporized from the surface
of the companion if their velocities exceed the escape velocity.
So, the method used in this paper is very simple. To examine
whether our method is reasonable or not, we use the same analytic
method in this paper to calculate the models in M00. We
re-calculated the model in Li \& van den Heuvel (\cite{LI97}),
using their method to get the SG model used in M00, and calculated
the stripped mass using our analytic method, which is shown by a
triangular point in Fig. \ref{Fig2}. Here, the kinetic energy of
the supernova ejecta is also from the W7 model of Nomoto et al.
(\cite{NTY84}) as used in M00. The stripped mass from our SG model
is smaller than that of M00, but the difference is not very
significant. We also calculate a 1 $M_{\rm \odot}$ solar model
used in M00 by Eggleton's stellar evolution code, and calculate
the stripped mass from this MS model using the same analytic
method in this paper. $a/R_{\rm 2}$ is changed according to M00,
not from Eggleton's equation (Eggleton \cite{EGG83}). The results
are shown by filled squares in Fig. \ref{Fig2}. A similar linear
relation between log($\delta M$) and log($a/R_{\rm 2}$) is gained
as indicated by M00. However, the stripped mass in our model is
smaller than that in M00 for small $a/R_{\rm 2}$ and larger than
that in M00 for large $a/R_{\rm 2}$, which is derived from our
simple method. Since the conservation of linear momentum and the
completely inelastic collision are applied, and the shock induced
in the secondary envelope by the impact of the ejecta is not
considered, the effect of ablation induced by the shock is not
considered. For the simple method, most of the energies which
should be used to heat the secondary envelope and to vaporize the
material in the envelope are lost with the stripped material for
small $a/R_{\rm 2}$, while for large $a/R_{\rm 2}$, a part of
energy which should heat the secondary envelope but were not used
to strip the material are collected to strip the material from the
surface of the companion in our model. Although the stripped mass
in our models is different from that of M00, our method can give a
similar trend to M00. We also use the same model as M00 and method
in this paper to calculate the kick velocity of the companion. The
results are shown in Fig. \ref{Fig3}. The difference between our
results and that of M00 is very small for all the models. This is
a natural result since the kick velocity is mainly decided by the
collision section of the companion for a given companion model. We
also gain a similar linear relation between log($V_{\rm kick}$)
and log($a/R_{\rm 2}$) to that indicated by M00. Then, although
the stripped mass is different from that of M00, the kick velocity
may be correct. Since we only want to discuss the trend of the
effect of some initial parameters, in this context, it is not
unreasonable for our method to do this. However, a fact must be
emphasized that since log($a/R_{\rm 2}$) concentrates in the range
of (0.35-0.5) in our models, the stripped mass in our models
should be taken as a lower limit for a real case, especially for
MS models.
\subsection{results}\label{subs:2.3}
  The stripped mass $\delta M$ and the ratio of $\delta M$ to the companion mass $M_{\rm 2}^{\rm
SN}$ at explosion are presented in panels (a) and (b) of Fig.
\ref{Fig4}. Although there is no obvious difference between the MS
and HG companions, the results seem to be divided into two groups
based on the mass transfer stage. For a given $a/R_{\rm 2}$, the
stripped mass $\delta M$ of the MS models is always smaller than
that of HG ones. Also, $\delta M/M_{\rm 2}^{\rm SN}$ of MS models
is always slightly smaller than that of HG ones at a certain
$M_{\rm 2}^{\rm SN}$. These differences are derived from the
different structure of the companions. If the mass transfer begins
as the companion is in tjhe HG, the companion has a denser core
and a more expanded envelope than the MS companion. Then, the
binding energy is smaller and the material in the envelope is
easier to strip off.

The kick velocities of the companions are shown in panel (d) of
Fig. \ref{Fig4}. $V_{\rm kick}$ is low and has little influence on
the spatial velocity of the companion, which is consistent with
the numerical simulation (Fryxell \& Arnett \cite{FRY81}; M00). In
Fig. \ref{Fig4}, we see that $V_{\rm kick}$ is relevant to the
mass-transfer stage: $V_{\rm kick}$ of a HG companion is always
smaller than that of a MS one at a certain $M_{\rm 2}^{\rm SN}$.
However, it is difficult to tell when the mass transfer begins
only according to a given kick velocity. There seems to exist a
peak value at a position of $M_{\rm 2}^{\rm SN}\simeq 1.0
M_{\odot}$. More calculation is needed to test whether this peak
value is real or not.

After the impact, the companion accretes a part of the ejecta and
will be puffed up, and its luminosity will increase sharply. At
the same time, the hydrogen-burning quenches because of its lower
central temperature and density and the companion is similar to a
pre-main-sequence star (M00). Because it is difficult to estimate
the thermal time scale of the companions in this situation, we
simply assume that the time scale for the companion to recover its
thermal equilibrium is $10^{\rm 4}$ yr (M00) for all of the
models.  Note however the thermal timescale actually depends on
the properties of the companion's envelope before SN Ia explosion
(Podsiadlowski \cite{POD03}), and our assumption oversimplifies
the problem. According to virial theorem, we assume that half of
the kinetic energy of the ejecta accreted by the companion is
radiated by photon energy. The companion's luminosity is estimated
via the half of the kinetic energy being divided by $10^{\rm 4}$
yr. The results are shown in panel (c) of Fig. \ref{Fig4}, which
are well consistent with the numerical simulation of M00. The
luminosity depends slightly on the mass transfer stage. For a
given $a/R_{\rm 2}$, the luminosity of the MS models is slightly
larger than that of the HG ones. The luminosity decreases with
$a/R_{\rm 2}$, which is a natural result since the collision
section of the companion decreases with $a/R_{\rm 2}$. Because of
the rough estimation of the thermal time scale here, we do not
discuss the relation between the luminosity and the initial
parameters of the binary system.

No obvious evidence shows that the stripped mass, $\delta M$, the
ratio of the stripped mass to companion mass, $\delta M/M_{\rm
2}^{\rm SN}$, and the kick velocity, $V_{\rm kick}$, depend
significantly on the initial metallicity and the initial WD mass.
The initial companion mass $M_{\rm 2}^{\rm i}$ and the initial
orbital period $P^{\rm i}$ affect the final result as shown in
Fig. \ref{Fig5}. In that figure, we see that both $\delta M$ and
$\delta M/M_{\rm 2}^{\rm SN}$ increase with $M_{\rm 2}^{\rm i}$
and $P^{\rm i}$, and $V_{\rm kick}$ increases with $M_{\rm 2}^{\rm
i}$ while it decreases with $P^{\rm i}$. These facts are relevant
to the evolutionary degree of the companion at explosion -- a more
massive star evolves more quickly and has a less compact envelope
at a given stage. Meanwhile, larger $P^{\rm i}$ results in the
companion being further away from zero age main sequence (ZAMS)
and the companion also has an less compact envelope at explosion.
The matter in this less compact envelope will be stripped off more
easily. For the same evolutionary reason, a larger $M_{\rm 2}$ and
a smaller $P^{\rm i}$ result in a larger collision section.

The $V_{\rm kick}$ is not high compared to the orbital velocity,
$V_{\rm orb}$. The ratios of $V_{\rm kick}/V_{\rm orb}$ locate in
the rang of 0.25 to 0.5. The spatial velocity, $V_{\rm 2}^{\rm
SN}=\sqrt{V_{\rm kick}^{\rm 2}+V_{\rm orb}^{\rm 2}}$, of the
companion after an explosion is mainly decided by the initial
parameters of the progenitor, except for metallicity. The spatial
velocity increases with the companion mass and decreases with the
WD mass and the period, which are natural results of binary
evolution. The spatial velocity ranges from 120 km/s to 200 km/s
and the velocity of star G (about 136 km/s) is located in this
range.

\section{Discussion and Conclusion}\label{sect:3}
\begin{figure}
   \vspace{2mm}
   \begin{center}
\includegraphics[width=3in, angle=-90]{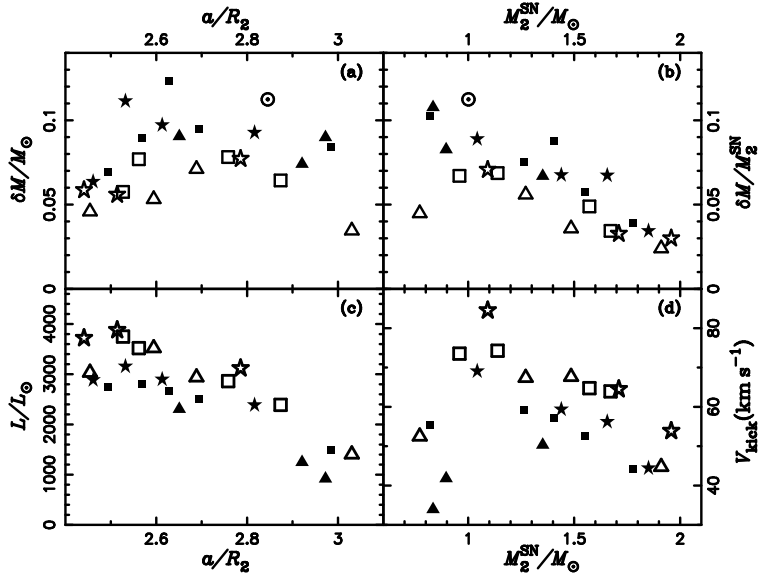}
   \caption{In panels (a) and (c), stripped mass, $\delta M$, and luminosity of companions, $L$,
   are shown as functions of the ratio of
   separation to the radius of companions, $a/R_{\rm 2}$. Panels (b) and (d)
   show the
   ratio of stripped
   mass to companion mass, $\delta M/M_{\rm 2}^{\rm SN}$, and the kick velocity, $V_{\rm kick}$, vs the
   companion mass at the moment of explosion.
   Triangles, squares and pentacles denote the cases
   for Z=0.01, 0.02 and 0.03, respectively. Filled symbols denote that mass transfer onsets at Hertzsprung gap and hollow
   symbols denote that mass transfer onsets at main sequence. Solar symbols are the results from
   a polytropic stellar model of 1 $M_{\odot}$.}
   \label{Fig4}
   \end{center}
\end{figure}

\begin{figure}
   \vspace{2mm}
   \begin{center}
\includegraphics[width=3in, angle=-90]{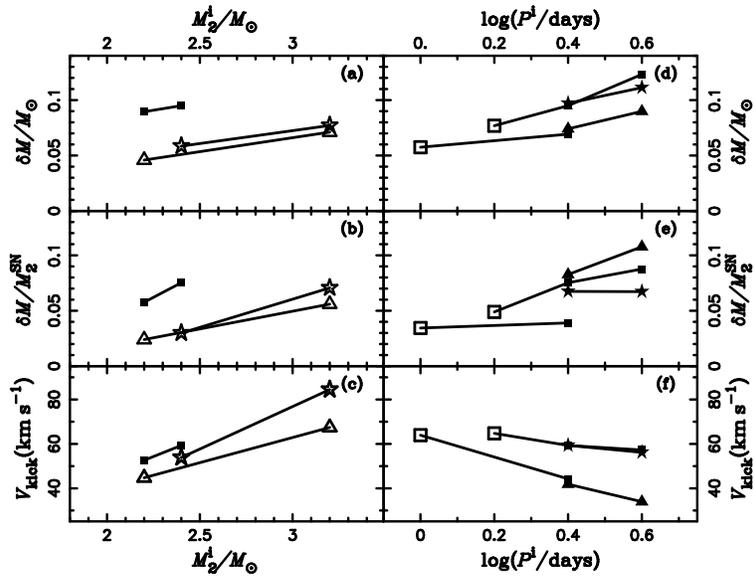}
   \caption{ Stripped mass, $\delta M$,
   ratio of the stripped
   mass to companion mass, $\delta M/M_{\rm 2}^{\rm SN}$, and kick velocity of the companion, $V_{\rm
   kick}$, vs the initial companion mass, $M_{\rm 2}^{\rm i}$, and orbital period, log$(P^{\rm i}\rm /day)$. The
   points linked by lines have same initial parameters excepting
   abscissas.
   The symbols are the same as in Fig. \ref{Fig5}}
   \label{Fig5}
   \end{center}
\end{figure}

By numerical simulation, Marietta et al. (\cite{MAR00}) performed
an excellent detailed study of the impact of a supernova's ejecta
on its companion. The study shows that there is no difference for
the stripped mass $\delta M$ and for the ratio of $\delta M/M_{\rm
2}^{\rm SN}$ between MS companions and SG companions. The simple
analytic solution in this paper is similar to the result of
numerical simulations. However, there seem to be two groups based
on the mass-transfer stage -- both $\delta M$ and $\delta M/M_{\rm
2}^{\rm SN}$ from SG models are always larger than those from MS
models at certain conditions, i.e. at a fixed $a/R_{\rm 2}$ or
$M_{\rm 2}^{\rm SN}$. This fact indicates that the companion
structure is important to discriminate $\delta M$ and $\delta
M/M_{\rm 2}^{\rm SN}$ during SNe Ia explosions. It is possible
that the process of mass transfer has a significant effect on the
final result. We did not find the linear relation between
log$(\delta M)$ and log$(a/R_{\rm 2})$ or between log$(V_{\rm
kick})$ and log$(a/R_{\rm 2})$ in Fig. \ref{Fig4} as given by M00.
This result is also relevant to the choice of the companion model
and this is because Marietta et al. (\cite{MAR00}) simply changed
$a/R_{\rm 2}$ for the same companion model to examine the effect
of $a/R_{\rm 2}$. The importance of the companion structure is
also verified by the fact that we can gain a similar relation to
M00 by adopting a similar assumption of M00.

A caveat must be emphasized that the kinetic energy of the
supernova ejecta translates into the thermal energy of the
companion envelope and a part of the material in the envelope is
heated and is vaporized to escape from the companion, which may
increase the stripped mass (Fryxell \& Arnett \cite{FRY81}; M00).
We do not consider this effect in the analytic solution although
this effect may affect the stripped mass significantly (Fryxell \&
Arnett \cite{FRY81}; M00). Therefore, as discussed in subsection
\ref{subs:2.2}, $\delta M$ in our models is only a lower limit.
Note that our analytic method may have oversimplified the physics
of the interaction between the ejecta and the companion star, e.g.
we did not calculate the effect of the shock formed between the
ejecta and the companion star. In this context, our results may be
taken as a qualitative one giving the trend of real case.

  However, Leonard (private communication 2007) showed the amount of
  the stripped mass may be less than 0.01 $M_{\odot}$ as derived from observation, although this result relies on
  the model of M00. The contradiction between the
  observation of Leonard and the prediction of M00 might be from
  the companion model used by M00, especially the effect of the mass transfer on the
  structure of the companion. For a realistic case, before SN Ia explodes, most material
  in the companion's envelope has transferred onto the CO WD. At the same time, the radius
  of the companion decreases (Langer et al. \cite{LAN00}). These
  facts make the companion more compact than that of a star with the same
  mass while without mass transfer, such as a solar model used in M00. It is more difficult to
  strip material from the envelope of a compact
  companion and the stripped mass in M00 should therefor be lower.
  The result that the stripped mass in this paper is lower than that in M00
  might go in the right direction, showing
  the importance of the companion models.

We do not find the dependence of the $\delta M$,  $\delta M/M_{\rm
2}^{\rm SN}$ and $V_{\rm kick}$ on the metallicity. However, the
result depends on an assumption that the mass of the evaporated
material is independent of the metallicity. Since there is not any
numerical simulation or analytic method to test this assumption,
we have no way to estimate the effect of the metallicity on the
evaporation by our simple method. Chugai (\cite{CHU86}) showed
that for the given explosion energy of a SN Ia, the mass of the
evaporated material is proportional to $(\rho a^{\rm 3})^{\rm
-0.5}$, where $\rho$ is the density of the companion's envelope at
the explosion and $a$ is the orbital separation at the explosion.
Increasing $(\rho a^{\rm 3})^{\rm -0.5}$ leads more mass
evaporated. According to our calculation, there is no systemic
effect of the metallicity on $\rho a^{\rm 3}$ and $\rho a^{\rm 3}$
is mainly decided by the mass transfer. Then, it is plausible that
there is no correlation between the metallicity and the $\delta
M$, $\delta M/M_{\rm 2}^{\rm SN}$, $V_{\rm kick}$.

The effect of the kinetic energy of supernova ejecta can be
examined by changing the kinetic energy. If $E_{\rm k}=1.5\times
10^{\rm 51}$ erg, which corresponds to the upper limit of the
kinetic energy of normal SNe Ia (Gamezo et al. \cite{GAM03}),
$\delta M$ increases by about $0.01 M_{\odot}$ compared with that
of $E_{\rm k}=1.0\times 10^{\rm 51}$ erg, the luminosity of the
companion increase by about 20\% to 30\% and the kick velocity
also increases slightly. An interesting phenomenon is that there
is a maximal spatial velocity  at infinity for the stripped
material and  this maximal velocity only depends on the kinetic
energy of the supernova ejecta. It is in the range of 8000 km/s to
9500 km/s for an $E_{\rm k}$ of $1.0\times 10^{\rm 51}$ erg to
$1.5\times 10^{\rm 51}$ erg. However, almost for all the models,
the special velocities of a half of the stripped materials are
less than 1100 km/s. This is roughly consistent with the numerical
simulation (820 km/s and 890 km/s for MS and SG models,
respectively (M00))

By the same analytic method, we use a polytropic stellar model of
1 $M_{\odot}$ to examine the influence of companion structure on
the results. These results are plotted in Fig. \ref{Fig4} as a
solar symbol. For a given condition, $\delta M$ and $\delta
M/M_{\rm 2}^{\rm SN}$ are much larger than that of our MS models
and even larger than that of our HG models. $L$ and $V_{\rm kick}$
increase by a factor of 2-9 and 3-8, respectively. These
differences indicate that the values of $\delta M$, $L$ and
$V_{\rm kick}$ are overestimated by using a polytropic stellar
model. Note the fact that we can gain similar results to M00 by
using models similar to those of M00, especially for kick
velocity. So the influence of companion structure is very
important. The different structures between the MS models and the
SG models result in different stripped mass, luminosity and kick
velocity. Then, the difference between our results and that in M00
might be from the different structure of the companion.

Star G is likely to be the companion of Tycho's supernova and it
has a lower spatial velocity and luminosity compared to
theoretical predictions (Canal et al. \cite{CAN01}; Marietta et
al. \cite{MAR00}). Our model may naturally interpret the spatial
velocity of Star G, while the luminosity of Star G is lower than
the prediction of our model and than that of the numerical
simulation of M00 (by about 3 orders of magnitude). Although this
result may be partly from our approximation of the thermal
timescale of the companion, it may still reflect the fact to some
extent. Podsiadlowski (\cite{POD03}) showed that if the energy
injected into the companion's envelope is larger than $10^{\rm
47}$erg, the luminosity of the companion after $10^{\rm 3}$yr is
higher than that of Star G by  at least one order of magnitude.
Noting that the energy injected into the companion's envelope in
all of our models is much larger than $10^{\rm 47}$erg and
considering that the time since Tycho supernova (SN 1572) exploded
is less than 500 yr, we suggest that an energy-loss mechanism
might be needed to explain the low luminosity of Star G. Much
effort is needed to solve this problem.

\begin{table*}
  \caption{Initial parameters, i.e. metallicity (Column 2), WD
mass (Column 3), companion mass (Column 4) and orbital period
(Column 5) for our binary system model. The stage when mass
transfer begin is shown in Column 6. }\label{Tab:1}
  \begin{center}
    \begin{tabular}{cccccc}
      \hline
       $N_{\rm model}$  & $Z^{\rm i}$ & $M^{\rm i}_{\rm WD}/M_{\odot}$ & $M^{\rm i}_{\rm 2}/M_{\odot}$ & log($P^{\rm i}$/day) & Onset \\
      \hline
        1 & 0.01 & 1.00 & 2.40 & 0.60 & HG\\
        2 & 0.01 & 1.00 & 2.40 & 0.40 & MS\\
        3 & 0.01 & 1.00 & 2.20 & 0.00 & MS\\
        4 & 0.01 & 1.10 & 3.20 & 0.20 & HG\\
        5 & 0.01 & 1.10 & 2.40 & 0.40 & HG\\
        6 & 0.01 & 1.20 & 3.20 & 0.20 & MS\\
        7 & 0.01 & 1.20 & 2.20 & 0.20 & MS\\
        8 & 0.02 & 0.75 & 2.00 & 0.20 & MS\\
        9 & 0.02 & 0.80 & 2.20 & 0.40 & HG\\
        10 & 0.02 & 1.00 & 2.20 & 0.40 & HG\\
        11 & 0.02 & 1.00 & 2.20 & 0.00 & MS\\
        12 & 0.02 & 1.00 & 2.40 & 0.20 & MS\\
        13 & 0.02 & 1.00 & 2.40 & 0.40 & HG\\
        14 & 0.02 & 1.00 & 2.40 & 0.60 & HG\\
        15 & 0.02 & 1.10 & 3.20 & 0.20 & MS\\
        16 & 0.02 & 1.10 & 2.20 & 0.40 & HG\\
        17 & 0.03 & 0.80 & 2.20 & 0.40 & HG\\
        18 & 0.03 & 1.00 & 2.20 & 0.00 & MS\\
        19 & 0.03 & 1.00 & 2.40 & 0.40 & HG\\
        20 & 0.03 & 1.00 & 2.40 & 0.60 & HG\\
        21 & 0.03 & 1.10 & 2.40 & 0.20 & MS\\
        22 & 0.03 & 1.10 & 3.20 & 0.20 & MS\\
        23 & 0.03 & 1.10 & 2.20 & 0.40 & HG\\
      \hline
    \end{tabular}
  \end{center}
\end{table*}

\end{document}